\documentclass[a4paper,11pt]{article}
\usepackage{authblk}
\usepackage{amsmath,amsthm,amssymb}
\usepackage[latin1]{inputenc}
\usepackage[swedish,english]{babel}
\usepackage{ifpdf}
\ifpdf
  \usepackage[pdftex]{graphicx}
  \usepackage{epstopdf}
\else
  \usepackage[dvips]{graphicx}
\fi
\usepackage{sidecap}
\usepackage[pdftitle={Stochastic simulation of growing patterns},
  pdfauthor={Stefan Engblom},
  pdffitwindow=true,
  breaklinks=true,
  colorlinks=true,
  urlcolor=blue,
  linkcolor=red,
  citecolor=red,
  anchorcolor=red]{hyperref}
\usepackage{algorithm}
\usepackage{algorithmic}
\usepackage[numbers,sort]{natbib}

\numberwithin{equation}{section}
\numberwithin{table}{section}
\numberwithin{figure}{section}
\numberwithin{algorithm}{section}

\theoremstyle{plain}

\theoremstyle{definition}

\theoremstyle{remark}

\newcommand{\ave}[1]{\langle #1 \rangle}
\newcommand{\Vol}{\Omega}

\newcommand{\Nvoxels}{N_{\mbox{{\tiny vox}}}}

\newcommand{\review}[1]{#1}

\title{Stochastic simulation of pattern formation in growing tissue: a
  multilevel approach}

\author[1]{Stefan Engblom\thanks{URL:
    \url{http://user.it.uu.se/~stefane}, telephone +46-18-471 27 54,
    fax +46-18-51 19 25.}}

\affil[1]{{\footnotesize Division of Scientific Computing, Department
    of Information Technology, Uppsala University, SE-751 05 Uppsala,
    Sweden. E-mail:
    \href{mailto:stefane@it.uu.se}{stefane@it.uu.se}.}}

\begin{document}

\selectlanguage{english}

\maketitle

\begin{abstract}
  We take up the challenge of designing realistic computational models
  of large interacting cell populations. The goal is essentially to
  bring Gillespie's celebrated stochastic methodology to the level of
  an interacting population of cells. Specifically, we are interested
  in how the gold standard of single cell computational modeling, here
  taken to be spatial stochastic reaction-diffusion models, may be
  efficiently coupled with a similar approach at the cell population
  level.

  Concretely, we target a recently proposed set of pathways for
  pattern formation involving Notch-Delta signaling mechanisms. These
  involve cell-to-cell communication as mediated both via direct
  membrane contact sites as well as via cellular protrusions. We
  explain how to simulate the process in growing tissue using a
  multilevel approach and we discuss implications for future
  development of the associated computational methods.


%% 60-XX 	Probability theory and stochastic processes
%% 60Jxx 	Markov processes
%% 60J22   	Computational methods in Markov chains
%% 60J27   	Continuous-time Markov processes on discrete state spaces
%% 60J28   	Applications of continuous-time Markov processes on discrete 
%%                   state spaces

%% 65-XX 	Numerical analysis
%% 65Cxx 	Probabilistic methods, simulation and stochastic differential 
%%                   equations
%% 65C40   	Computational Markov chains

%% 92-XX 	Biology and other natural sciences
%% 92-08   	Computational methods

\end{abstract}

\selectlanguage{english}

\maketitle

%**************************************************************************

\section{Introduction}

% challenge
An important challenge in computational cell biology is to study the
emergent behavior of single-cell pathways at the scale of a large
interacting cell population. In this paper we tackle this challenge
by, in essence, attempting to generalize Gillespie's stochastic
simulation methodology to the level of the multicellular
environment. In order to do so, clearly, the modeling physics of the
extracellular space, of the cell population, and of the single cells
need to be prescribed. A suitable computational methodology should
additionally allow for cell-to-cell signaling in a flexible and
general way. There are several possible interesting applications for
such a kind of modeling framework; regulating processes in embryonic
development, angiogenesis, neurogenesis, wound healing, and tumor
growth, to mention just a few.

%% Most first-principle formulations are both multiphysics and
%% multiscale in nature, and, moreover, stochastic formulations are
%% often preferred to better capture the effects of intrinsic noise.

% intro to Notch, why suitable as a target
In the interest of focusing our work around a concrete, yet fairly
demanding modeling situation, we pick as our target a specific set of
network models which involve single-cell pathways together with
non-trivial signaling between the individual cells. The Notch
signaling pathway is a highly conserved mechanism which is present in
most multicellular organisms \cite{notch_review}, ranging from, e.g.,
\textit{Drosophila} and \textit{C.~elegans} to mammals. Indeed, the
fundamental importance of Notch signaling made it an early target for
mathematical models \cite{delta_notch}, where feedback regulation
between neighbor cells was modeled. It has since been realized that
cell-to-cell signaling not only \review{is short range,} taking place
at direct junctional contact sites, but also is mediated via
\review{long range} cellular protrusions
\cite{filopodia}. Mathematical models including these effects have
recently been investigated \cite{mutualNotch, delta_notch_report} and
we choose a family of such models as the concrete target in this
paper.

% short intro to single cell modeling via the RDME
To be able to \review{realistically resolve the geometrical details}
of the single cell, unstructured meshes (e.g., triangularizations)
stand out as a ubiquitous tool. Also, an important part of Dan
Gillespie's heritage to computational biology is that noisy cellular
processes at the molecular level should be understood in a
\emph{stochastic} framework. These observations together suggest the
\emph{reaction-diffusion master equation} (RDME) over an unstructured
mesh \cite{master_spatial} and we shall regard it herein as a gold
standard in single cell modeling. \review{The RDME is based on first
  principles and is reasonably effective
  computationally. Additionally,} this description, or simplified
versions of it, has been successful at delivering important insights
for a range of cellular phenomena \cite{NSM,stochgeneexpression,
  fluctuation_limits, circadian}.

% short intro to on-lattice cell-based computational modeling
At the scale of a population of cells, cell-based computational
modeling is an \textit{in silico} approach to test hypotheses
concerning the contributions of various mechanisms to observed
macro-level behaviors. Examples of recent applications of cell-based
models include embryonic development \cite{Atwell:2015:MLM}, wound
healing \cite{Vermolen:2013:ASC, Ziraldo:2013:CMO}, and tumor growth
\cite{Naumov:2011:CAM}. The natural analogue of the RDME at the cell
population level is found in the class of \emph{on-lattice} cell-based
models. As in the RDME, space is here discretized in a grid of voxels
over which the cells are distributed. State update rules are then
formulated over this grid where signaling processes and factor
concentrations may be included via, e.g., differential equations
\cite{Robertson:2015:IOM}.

% approach here: RDME+DLCM
In this work we will focus on the novel on-lattice method proposed in
\cite{laplace_cellmech}, which is promising from a scaling point of
view, yet also is very expressive. The method is referred to as
\emph{discrete Laplacian cell mechanics} (DLCM), and is formed by
developing constitutive equations for the dynamics of the cell
population at a given discretization of space. The update rules are
stochastic and are established from global calculations. Importantly,
the simulations take place in continuous time, thus allowing for a
meaningful coupling to arbitrary continuous-time processes, including,
e.g., inter-cellular signals. In summary, we focus our attention to
single cell models described in the RDME framework and the main
contribution of the paper is to investigate the feasibility of the
two-level RDME-DLCM approach.

% overview
In the next section we work through the specific, but fairly general,
pattern forming mechanism we wish to study and we subsequently express
it within the RDME-DLCM computational framework. As will be
demonstrated this enables us to simulate a range of intriguing
patterns in an unprecedented detailed and bottom-up fashion. The paper
is concluded with a discussion of some ideas concerning possible
future developments of the presented computational methodology.

%**************************************************************************

\section{Models and methods}

Below we start by presenting the specific Notch pathway model we will
target in the paper. Throughout we consider a single non-dimensional
model, originating from an attempt to map to the situation of
explaining the organization of bristles on the \textit{Drosphila}
notum \cite{filopodia}. \review{Such patterns are remarkably precise
  and are therefore good model systems to study the genetic basis of
  pattern formation.} The model we decided to employ can be found in
\cite{delta_notch_report}. We make a slight extension of the model in
\S\ref{subsec:RDME} by bringing it into the spatial setting,
essentially by deciding on a system volume and settling for suitable
diffusion constants. In \S\ref{subsec:DLCM} we explain how to use the
methodology in \cite{laplace_cellmech} to efficiently simulate a
growing cell population. The two computational layers, i.e., the
single cell- and the cell population layer, are put together in
\S\ref{subsec:results} where we present a few selected simulation
results. In order to concentrate on the possibilities with the
computational framework we select spatial- and cell population
parameters rather freely, and we do not claim our resulting model to
map to any specific real-world scenario.

\subsection{Protrusion mediated Notch-Delta pattern formation}
\label{subsec:NDR}

An early attempt to mathematically explain pattern formation
mechanisms in tissue without resorting to the postulated existence of
\emph{morphogens}, i.e, as done early on by Turing
\cite{morphogenesis}, was based on lateral inhibition with feedback
\cite{delta_notch}. This mechanism takes place in between the
trans-membrane proteins Notch and Delta, respectively. In a
non-dimensional setting, with $(n_i,d_i)$ denoting the Notch- and Delta
concentrations within cell $i$, the original model has the form
\cite{delta_notch}
\begin{align}
  &\left. \begin{array}{rcl}
    \label{eq:delta_notch_ODE}
    n'_i &=& f(\ave{d}_i)-n_i \\
    d'_i &=& \mbox{const.} \times \left( g(n_i)-d_i \right) \\
  \end{array} \right\}
\end{align}
where $'$ denotes differentiation with respect to time and where
$\ave{d}_i$ denotes the incoming Delta signal, averaged from the cells
surrounding cell $i$. In \eqref{eq:delta_notch_ODE}, $f$ and $g$
denote monotonically increasing and decreasing functions of their
single argument, respectively.

Whereas the classical Notch-Delta model gives an alternating pattern
of 'black' (e.g., high Delta) and 'white' (low Delta), patterns in
Nature are often much more involved, e.g., with sparse dots, or spots,
stripes, and labyrinth-like patterns. In an attempt to explain the
dot-like pattern of the notum (dorsal portion of the thoracic segment)
of \textit{Drosophila}, communication of Delta via cellular
protrusions was added to the model in \cite{filopodia}. Later details
were added in \cite{mutualNotch, delta_notch_report}, including
differential weighting of the incoming signals and an addition of the
concentration of a Notch reporter molecule $r_i$. More specifically,
the model from \cite{delta_notch_report} reads
\begin{align}
  &\left. \begin{array}{rcl}
    \label{eq:NDR}
    n'_i &=& \beta_n-\frac{\ave{d_{in}}n_i}{k_t}-\frac{d_in_i}{k_c}-n_i \\
    d'_i &=& \beta_d\frac{1}{1+r_i^m}-\frac{d_i\ave{n_{in}}}{k_t}-
    \frac{d_in_i}{k_c}-d_i \\
    r'_i &=& \beta_r\frac{(\ave{d_{out}}n_i)^s}{k_{rs}+(\ave{d_{out}}n_i)^s}-r_i \\
  \end{array} \right\}
\end{align}
\review{In \eqref{eq:NDR}, Delta $d_i$ is down-regulated by the Notch
  reporter $r_i$, which in turn is up-regulated by Notch $n_i$ and,
  respectively, the outgoing Delta $\ave{d_{out}}$, as discussed
  below. The Hill coefficients in these regulations are taken to be $m
  = s = 2$ throughout the paper.}

\begin{SCfigure}
  \includegraphics[width=0.5\textwidth,trim = 1cm 0.5cm 0.5cm 0.5cm,clip =
    true]{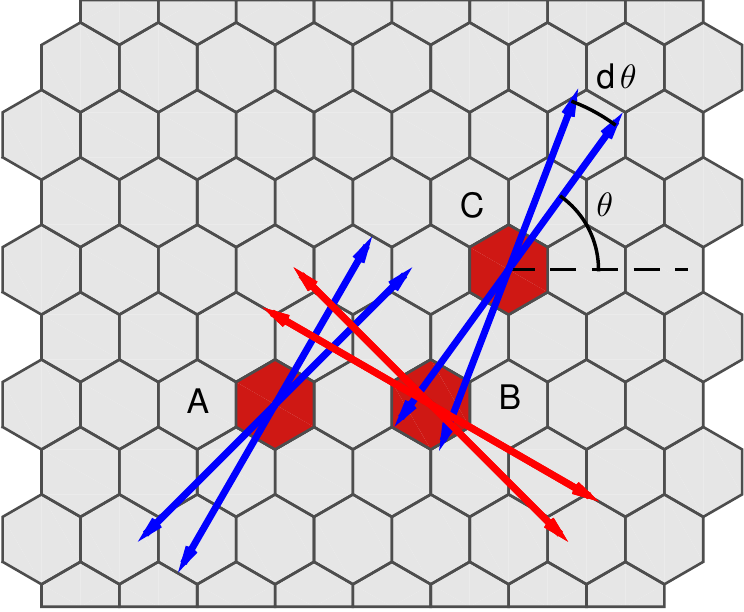}
  \caption{Signaling via protrusions: the symmetric contact $A
    \longleftrightarrow B$ is protrusion mediated, $C \longrightarrow
    B$ is protrusion-to-membrane, and $B \longrightarrow C$
    membrane-to-protrusion. A junctional contact is also possible
    between neighbor cells (not shown). In the running model of the
    paper, the first two types of contacts are understood to be
    protrusional (superscript $(b)$), while the two latter types are
    junctional (superscript $(a)$). \review{Protrusions are
      parameterized by the protrusion length $l$, direction $\theta$,
      and angular width $d\theta$.}}
  \label{fig:protrusions}
\end{SCfigure}

The existence of protrusional communication clearly implies more
cell-to-cell signaling possibilities
(Figure~\ref{fig:protrusions}). In \eqref{eq:NDR}, the amount of
incoming \review{and outgoing} Delta and Notch, respectively, is given
by
\begin{align}
  \label{eq:signals}
  &\left. \begin{array}{rcl}
    \ave{d_{in}} &=& w_a \ave{d}_i^{(a)}+w_b \ave{d}_i^{(b)} \\
    \ave{d_{out}} &=& q_a \ave{d}_i^{(a)}+q_b \ave{d}_i^{(b)} \\
    \ave{n_{in}} &=& w_a \ave{n}_i^{(a)}+w_b \ave{n}_i^{(b)}
  \end{array} \right\}
\end{align}
The superscript $(a)$ and $(b)$ denote the sum of the signal over all
cells making \emph{junctional} and, respectively, \emph{protrusional}
contact, see Figure~\ref{fig:protrusions}. \review{More concretely,
\begin{align}
  \ave{d}_i^{(a)} &:= \sum_{j \in J(i)} d_j, \qquad
  \ave{d}_i^{(b)} := \sum_{j \in P(i)} d_j,
\end{align}
in which $J(i)$ and $P(i)$, denote the set of junctional and
protrusional contacts for cell $i$.} A specific novelty with this
model is $\ave{d_{out}}$, the total amount of bound Delta that leads
to activation of the Notch receptor. Differential weighting of the
signals is achieved by assuming different constant weights
$[w_a,w_b,q_a,q_b]$ of the incoming signals.

To formulate a stochastic well-stirred interpretation of
\eqref{eq:NDR} we understand the concentrations $(n,d,r)$ in cell $i$
as absolute molecular counts $(N,D,R)$ at some fix system volume
$\Vol$. From \eqref{eq:NDR} we propose the transitions
\begin{align}
  \label{eq:NDR_SSA}
  &\left.\begin{array}{lll}
     \emptyset \xrightarrow{\beta_n \Vol} N & \quad 
     \emptyset \xrightarrow{\beta_d \Vol \, r_1} D & \quad 
     \emptyset \xrightarrow{\beta_r \Vol \, r_2} R \\
     N \xrightarrow{\ave{D_{in}}/(k_t\Vol)} \emptyset & \quad 
     D \xrightarrow{\ave{N_{in}}/(k_t\Vol)} \emptyset & \quad 
     N+D \xrightarrow{1/(k_c \Vol)} \emptyset \\
     N \xrightarrow{1} \emptyset & \quad 
     D \xrightarrow{1} \emptyset & \quad
     R \xrightarrow{1} \emptyset
   \end{array} \right\} \\
  \intertext{in terms of}
  r_1 &\equiv \frac{1}{1+(R/\Vol)^2}, \quad
  r_2 \equiv \frac{(\ave{D_{out}}N/\Vol^2)^2}{k_{rs}+(\ave{D_{out}}N/\Vol^2)^2}.
\end{align}

\begin{figure}
  \centering
  \includegraphics[width=0.24\textwidth]{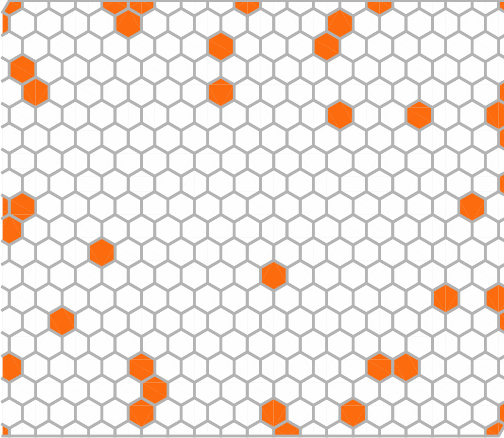}
  \includegraphics[width=0.24\textwidth]{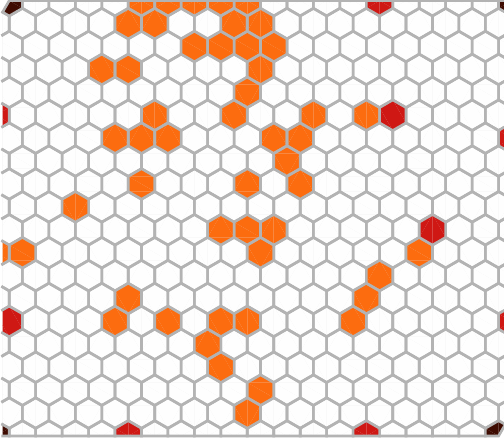}
  \includegraphics[width=0.24\textwidth]{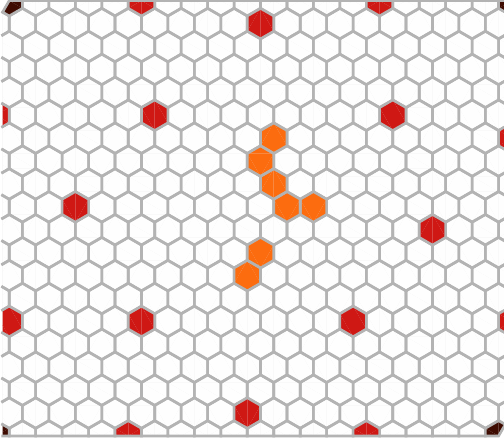}
  \includegraphics[width=0.24\textwidth]{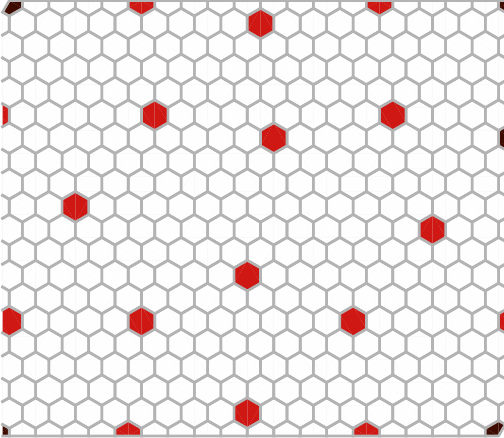}
  \caption{Stochastic pattern development over a static population of
    cells. Here the 0-dimensional interpretation of the model
    \eqref{eq:NDR_SSA} with system volume $\Vol = 400$ is employed and
    is simulated using Gillespie's Direct method. From left to right:
    time $t = [4,20,40,200]$. Color codes low (white) and high Delta
    (\review{orange, light- and dark brown}). Parameters are adopted
    from \cite{delta_notch_report}: $[\beta_n,\beta_d,\beta_r] =
    [100,500,3 \cdot 10^5]$, $[k_t,k_c,k_{rs}] = [2,0.5,10^7]$,
    protrusion length $= 3.5$ cell radii, \review{angular width $=
      2\pi$,} $[w_a,q_a,w_b,q_b] = [1,1,1,1]$.}
  \label{fig:staticSSA}
\end{figure}

We test this model over a a static hexagonal grid using Gillespie's
Direct method for the simulation (Figure~\ref{fig:staticSSA}). To
practically evaluate the various incoming signals $\ave{D_{in}}$,
$\ave{N_{in}}$, and $\ave{D_{out}}$, we settle for a small time-step
$d\tau$ and make the approximation that the signals \eqref{eq:signals}
remain constant in the interval of time $[t,t+d\tau]$\review{, in line
  with, e.g., the approach taken in \cite{bridging_the_gap}}. The time
interval was here chosen in a quite conservative way such that a
forward Euler step \review{of the original ODE-model \eqref{eq:NDR}}
would imply a 5\% change of state in a norm-wise sense,
\begin{align}
  \label{eq:dtau}
  d\tau &= 0.05 \times \frac{\|x\|}{\|f(x)\|},
\end{align}
where $x = [n,d,r] \review{= [N,D,R]/\Vol}$ is the concentration
vector for the whole cell population, and where $f(\cdot)$ is the
right-hand side of \eqref{eq:NDR}.

Simulation results for this model when starting from a random initial
configuration are summarized in Figure~\ref{fig:staticSSA} at a system
volume $\Vol = 400$. We now proceed to make an immediate spatial
extension of this model.

\subsection{Spatial stochastic reaction-diffusion models of single cells}
\label{subsec:RDME}

Living cells are inherently inhomogeneous objects and the assumption
of well-stirredness can rightly be questioned \cite{Dobrzynski,
  NSM}. The reaction-diffusion master equation (RDME) attempts to
strike a balance between accuracy and computational efficiency
\cite{Gardiner}. Here the domain under consideration is discretized in
small enough compartments, or \emph{voxels}, such that diffusion is
enough to regard each voxel as well-stirred. Diffusion in between
voxels are handled as a special set of reactions with rates obtained
so as to match with macroscopic diffusion properties
\cite{master_spatial}. An efficient algorithm for spatial stochastic
simulation is the Next subvolume method (NSM) \cite{NSM}, which can be
thought of as a blend of Gillespie's Direct method with the Next
reaction method. The algorithm is summarized in
Appendix~\ref{app:algorithms}.

We like to regard the RDME as a kind of ``gold standard'' in single
cell modeling. Although it is possible to make more accurate
computational models in the sense of bringing in more physical
details, say at the level of single molecules
\review{\cite{Andrews_single_molecule}}, this comes at large
computational costs. There is also the issue with uncertainties in
rate parameters, and the risk of over-modeling in many situations of
practical biological interest.

\begin{figure}
  \includegraphics{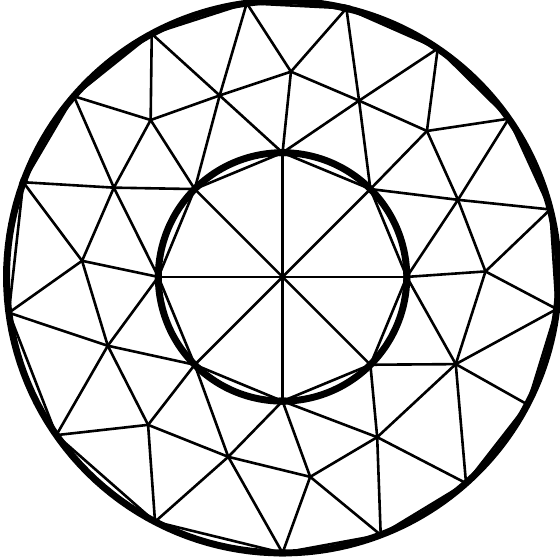}
  \includegraphics{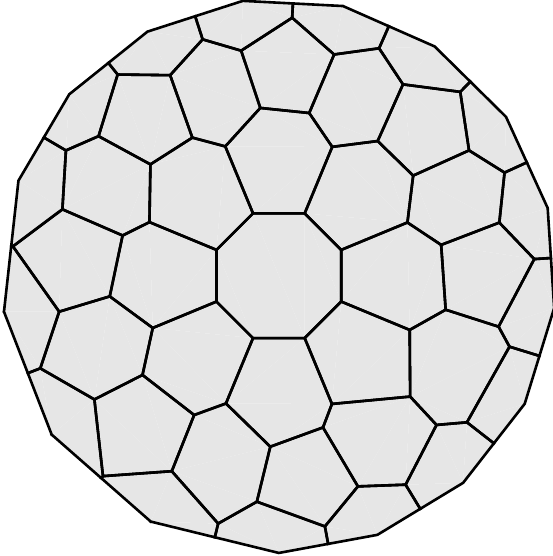}
  \caption{Single cell discretization mesh used in the running
    model. \textit{Left:} triangularization of a basic two-dimensional
    cell geometry consisting of a cytoplasm and a nuclei,
    \textit{right:} the associated dual mesh consisting of the
    computational compartments (voxels).}
  \label{fig:RDME}
\end{figure}

To illustrate the single cell-population level approach we have in
mind we make an immediate version of the pathway model
\eqref{eq:NDR_SSA} as follows. As a single cell discretization we take
the triangularization depicted in Figure~\ref{fig:RDME} which consists
of a modest number of 40 voxels. We make no particular distinction of
the membrane, the cytoplasm, or the nuclei, but allow all reactions in
\eqref{eq:NDR_SSA} to take place in all of the voxels. We let the
geometry be of total volume $\Vol = 400$ and use the scalar diffusion
constant $1/\Vol$ across the whole cell geometry and for all species
$[N,D,R]$. Although there are clearly many potential improvements to
this basic model it will serve as an interesting load case to our
simulation approach. We thus have to postpone for another occasion the
interesting quest for additional modeling realism including, e.g.,
nuclei- and membrane specific transitions.

\subsection{Stochastic simulation of growing cell populations}
\label{subsec:DLCM}

Given the relative efficiency of the RDME approach one can wonder if
not a similar idea could be useful at the cell population
level. Unlike the various molecules inside the living cell, however,
cells in multicellular structures do not generally diffuse around
freely. Instead, cells may actively crawl, adhere to other cells, and
are pushed into position. An RDME-like framework for this situation
was recently developed and we now briefly review this idea
\cite{laplace_cellmech}.

\begin{SCfigure}
  \centering
  \includegraphics[width = 0.5\textwidth,clip=true,trim=0.5cm 0.5cm 0 0]{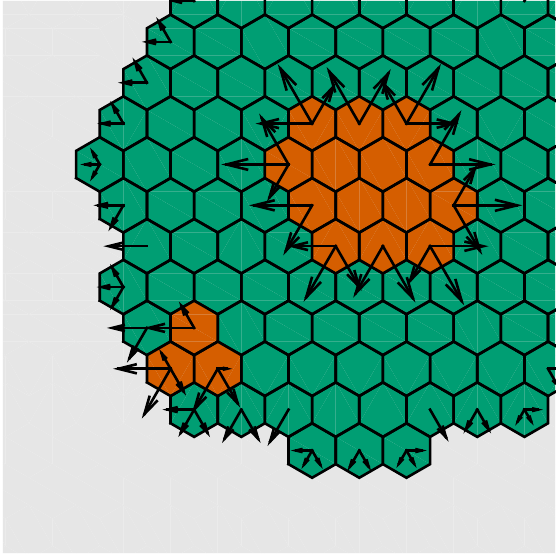}
  \caption{Schematic illustration of the DLCM method (adapted from
    \cite{laplace_cellmech}). \textit{Green} voxels contain single
    cells and \textit{red} voxels contain two cells, giving rise to a
    cellular pressure. A discrete Laplace operator is employed to
    propagate this pressure, thus inducing a rate to move for the
    cells in the voxels as indicated by the arrows. Cells in boundary
    voxels may move into empty voxels and cells in doubly populated
    voxels may move into voxels containing fewer cells.}
  \label{fig:basic_schematics}
\end{SCfigure}

We assume a two- or three dimensional computational grid consisting of
voxels $(v_i)$, $i = 1,\ldots,\Nvoxels$
(Figure~\ref{fig:basic_schematics}). \review{At this level of
  description the individual cells are placed in the single voxels of
  a typically structured grid, e.g., squares or hexagonals, although
  unstructured grids are certainly also a possibility.} As in the RDME
it is fundamental that a consistent Laplace operator may be defined
over this grid; hence the name discrete Laplacian cell mechanics, or
DLCM.

The voxels are either empty or may contain a certain number of cells,
here taken to be either 1 or 2. If the number of cells in all voxels
are either 0 or 1, i.e., below the carrying capacity, then the system
is in equilibrium given that there are no other active processes. If
the number of cells in one or more voxels is 2, a cellular pressure is
exerted towards the neighbor voxels. This state is eventually changed
by an event where one of the cells moves into a neighboring voxel, and
then the pressure distribution changes. This process continues until,
possibly, the system relaxes into equilibrium.

What is then the physics for this cellular pressure which drives the
shape of the cell population? A detailed derivation is made in
\cite{laplace_cellmech}, but in short, the answer is that a suitable
physics is formed by letting the pressure be spread according to the
negative Laplacian, and with source terms for all over-occupied
voxels, see Figure~\ref{fig:basic_schematics}. Let $(u_i)$,
$i = 1,\ldots,\Nvoxels$ denote the number of cells in voxel $i$ and
let $(p_i)$ be the corresponding cellular pressure. Denote by
$\Omega_h$ the subset of voxels $v_i$ for which $u_i \not = 0$ and let
$\partial \Omega_h$ denote the discrete boundary; the set of
unpopulated voxels that share an edge with a voxel in $\Omega_h$. At
any instant in time $t$ we solve for the cellular pressure,
\begin{align}
  \label{eq:dlaplace}
  -L p  &= s(u), \quad i \in \Omega_h, \\
  \label{eq:hom_dirichlet}
  p_i &= 0, \quad i \in \partial \Omega_h,
\end{align}
in which $L$ is a consistent discretization of $\Delta$ over
$\Omega_h$ and where the source term is $s(u_i) = 0$ for $u_i \le 1$
and $s(u_i) = 1$ whenever $u_i = 2$. This normalization ensures that
$p = 0$ at equilibrium. \review{It is doable to rely on this set-up
  also for unstructured meshes by postulating that the cellular
  pressure is proportional to the difference in volume occupancy and
  voxel volume. However, there are biological specifics which should
  rightly be considered in this case, such as adhesion effects in
  voxels populated under their carrying capacity, and also details
  concerning the volume characteristics of the individual cells.}

The movements in the cell population are induced by a pressure
gradient between two voxels. Denote by $I(i \to j) = I_{ij}$ the
current from voxel $v_i$ to the neighbor voxel $v_j$. This current is
found by integrating the pressure gradient across the edge between the
two voxels,
\begin{align}
  I_{ij} &= -\int_{v_i \cap v_j} \nabla p(x)
  \, \cdot dS = \frac{e_{ij}}{d_{ij}} (p_i-p_j),
\end{align}
with $d_{ij}$ the distance between voxel centers and $e_{ij}$ the
common edge length. The \emph{rate} for the event that one cell moves
from voxel $i$ to $j$ is taken to be
\begin{align}
  \label{eq:rate}
  R(i \to j) = R_{ij} = D I_{ij},
\end{align}
where the conversion factor $D$ may depend on position and on the type
of movement. We take $D = 0$ for movements into voxels containing an
equal number of cells, thus limiting the cellular movements to less
crowded voxels only.

The implied simulation method is event-based and is directly based on
Gillespie's Direct method, see Appendix~\ref{app:algorithms},
Algorithm~\ref{alg:DLCM}. Chiefly, for any given state of the cell
population, the rates of all events are determined and the time and
kind of the next event is sampled. Until the time of this next event,
any other processes local to each voxel may be simulated. When the
event is processed a new cell population state is obtained and the
process starts anew.

\begin{figure}
  \centering
  \includegraphics[width=0.24\textwidth]{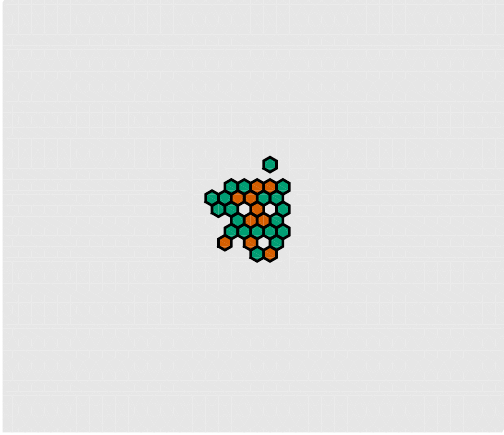}
  \includegraphics[width=0.24\textwidth]{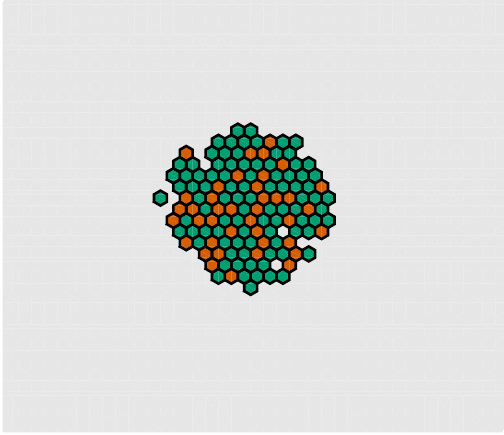}
  \includegraphics[width=0.24\textwidth]{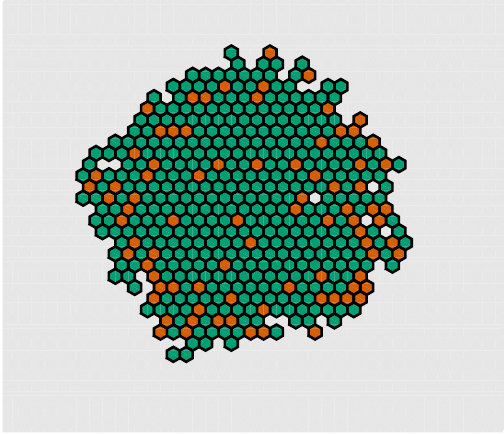}
  \includegraphics[width=0.24\textwidth]{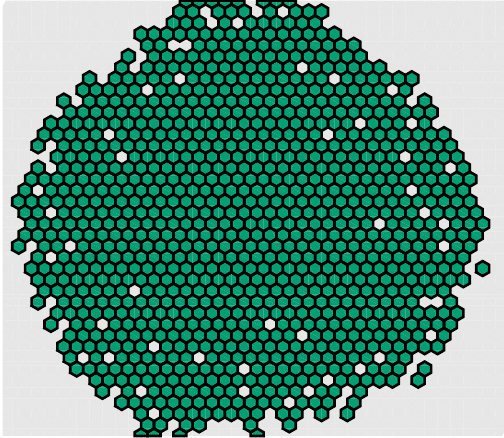}
  \caption{Growth of cell population. Nutrition is made available at
    the boundaries and diffuses through the population and is consumed
    by individual cells. From left to right: time $t =
    [14.4,21.8,31.9,197.9]$.}
  \label{fig:dynamicDLCM}
\end{figure}

We exemplify the process by growing a small population of 1000 cells,
starting from a single cell and allowing it to proliferate at a
certain rate provided it has enough concentration of `nutrition'. The
nutrition is distributed at the boundaries $\Omega_h$ of the cell
population and we let it diffuse by the Laplace operator. At any given
time, cells consume nutrition for their own metabolism and so this
scheme will favor the proliferation of cells near the boundary where
the nutrition concentration is the highest, see
Figure~\ref{fig:dynamicDLCM}. In the next section we proceed by
coupling this DLCM-layer growth process to the previously developed
RDME-layer description of the Notch-Delta pattern formation
mechanism. \review{Hence the fine RDME discretization as depicted in
  Figure~\ref{fig:RDME} is used to describe the physics of the
  individual cell, whereas the DLCM grid is used for the cell
  population.}

% *** DLCM-parameters
%% % diffusive pressure rate
%% Drate = 1;

%% % oxygen consumption rate per cell
%% gamma = 0.05;

%% % proliferation: rate and cutoff
%% lam = 0.5;
%% prolif_cutoff = 0.5;

\subsection{A range of Notch-Delta patterns in growing tissue}
\label{subsec:results}

\begin{figure}
  \includegraphics[width = 0.32\textwidth]{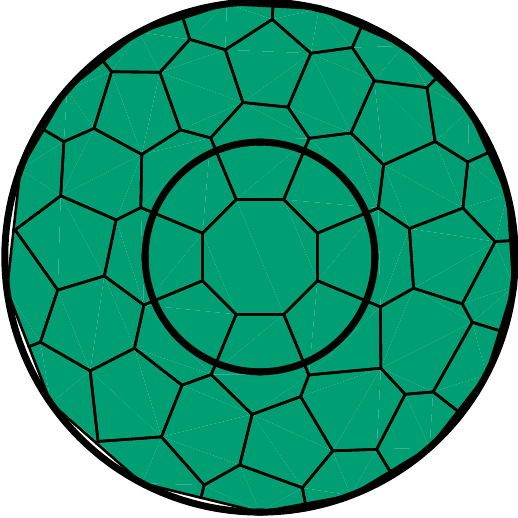}
  \includegraphics[width = 0.32\textwidth,clip = true,trim = 1cm 0.1cm 1cm 0.5cm]{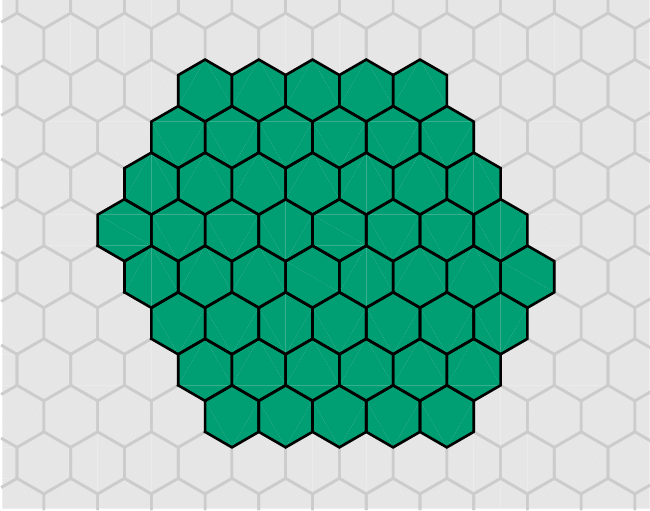}
  \includegraphics[width = 0.32\textwidth]{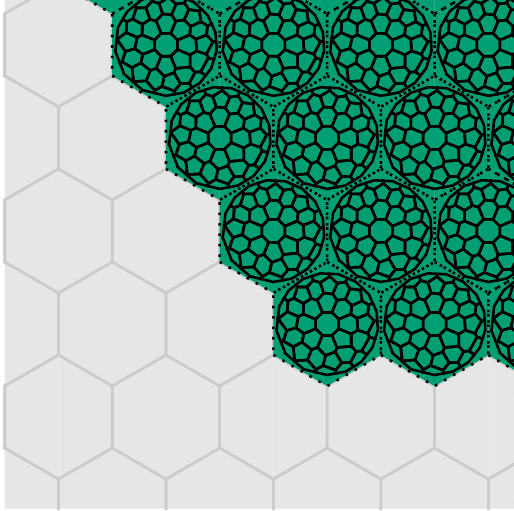}
  \caption{Inner-outer structure of the RDME-DLCM simulation
    approach. \textit{Left:} the single cell-model is simulated across
    all cells in the population displayed in the middle. The coupling
    between the cells which is required to capture, e.g., signaling
    processes, can be handled by a split-step time discretization
    strategy. \textit{Middle:} this continues until an event at the
    population layer is sampled. After executing this event and
    updating the internal states of the individual cells accordingly,
    the single cell model is evolved anew. \textit{Right:} the
    effective grid induced by this computational process can be
    understood as replicas of the single cell discretization.}
  \label{fig:inner_outer}
\end{figure}

In the present case the cellular growth process is independent from
the single-cell model and the two layers can be conveniently simulated
under a simple one-way coupling. That is, in a first run we simulate
the growth process and record all associated events separately. Next,
the RDME-layer model is simulated in continuous time and in between
all the recorded events, thus realizing the overall dynamics. This
coupling restriction is \review{only} used for convenience here: the
DLCM-simulation, cf.~Algorithm~\ref{alg:DLCM} in
Appendix~\ref{app:algorithms}, allows for a \review{completely}
general two-way coupling. However, in practice, it usually makes sense
to assume some kind of scale separation between the two layers
\review{\cite{multiscaleAgentCell}, such that the inner split-step
  $d\tau$ need not be much smaller than what is required to resolve
  the dynamics at the outer layer. Regardless of such an assumption,
  the split-step strategy can be expected to be strongly convergent in
  the stochastic sense \cite{jsdesplit}, although the split-step
  $d\tau$ might be severely restricted for accuracy reasons.}

\begin{algorithm}[H]{\small
  \caption{Details from the inner layer simulation of
    Algorithm~\ref{alg:DLCM}.}
  \label{alg:DLCMinner}
  \begin{algorithmic}

    \STATE{From Algorithm~\ref{alg:DLCM}: \textit{``Update the state
        of all cells with respect to any other continuous-time
        processes taking place in $[t,\tau)$, e.g., intracellular
        kinetics or cell-to-cell communication''}:}
    
    \WHILE{$t < \tau$}

    \STATE{Select a suitable time-discretization step $d\tau$, e.g.,
      according to \eqref{eq:dtau}.}

    \STATE{Simulate in time $[t,t+d\tau)$ and in parallel the
      single-cell RDME-model of all cells in all populated voxels
      $(v_i)$, $i = 1,\ldots,\Nvoxels$, e.g., using
      Algorithm~\ref{alg:NSM}.}

    \STATE{Update the cell-to-cell signals $\ave{D_{in}}$,
      $\ave{N_{in}}$, and $\ave{D_{out}}$, cf.~\eqref{eq:signals}.}

    \STATE{Set $t = t+d\tau$.}
    
    \ENDWHILE
  \end{algorithmic}}
\end{algorithm}

In Algorithm~\ref{alg:DLCMinner} we expand the details of the inner
layer simulation where the discretization in time chunks $[t,t+d\tau)$
is made explicit. It follows that the assumption made here is
essentially that the Notch-Delta dynamics takes place on a faster
time-scale than the growth process, a quite reasonable assumption in
this case. Without this assumption the simulation efficiency will
deteriorate whenever $d\tau$ for accuracy reasons has to be chosen
small.

\begin{figure}[H]
  \includegraphics[width=0.24\textwidth]{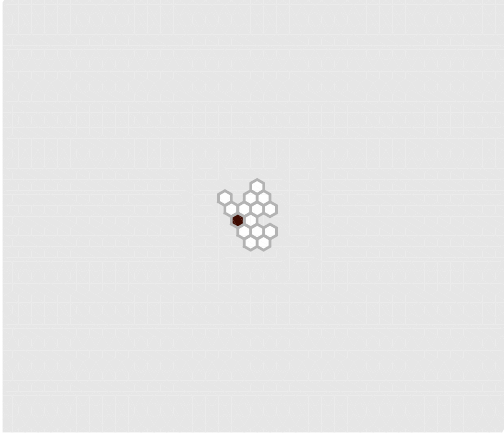}
  \includegraphics[width=0.24\textwidth]{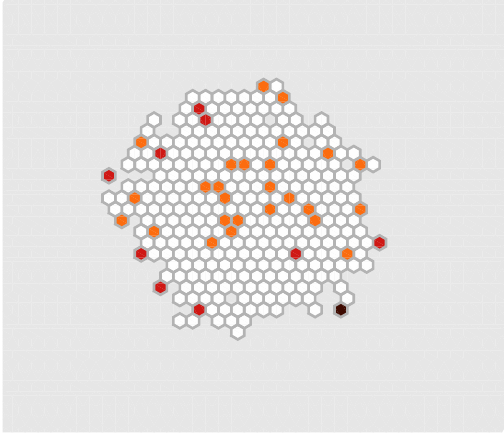}
  \includegraphics[width=0.24\textwidth]{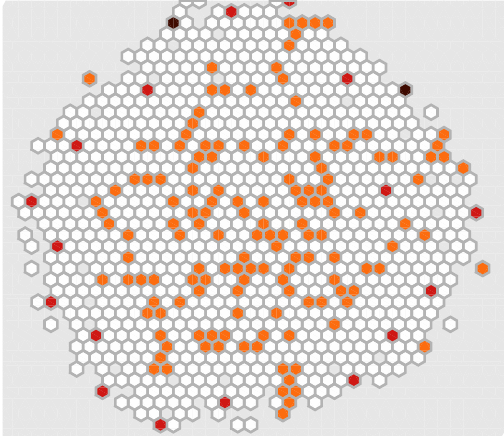}
  \includegraphics[width=0.24\textwidth]{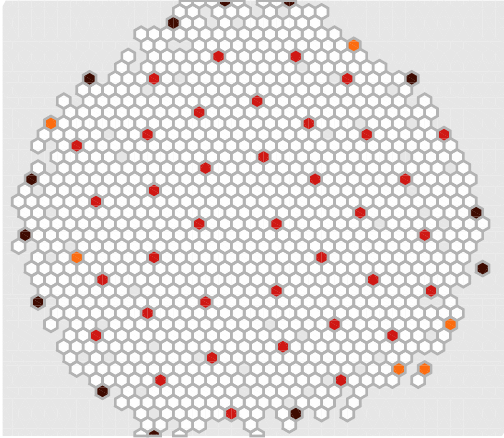}
  \caption{Notch-Delta-Reporter model in a growing domain. The RDME is
    used to describe the individual cells and the DLCM models the cell
    population growth. The parameters are as in
    Figure~\ref{fig:staticSSA} and \ref{fig:dynamicDLCM}.}
  \label{fig:dynamicRDME1}
\end{figure}

\begin{figure}
  \includegraphics[width=0.24\textwidth]{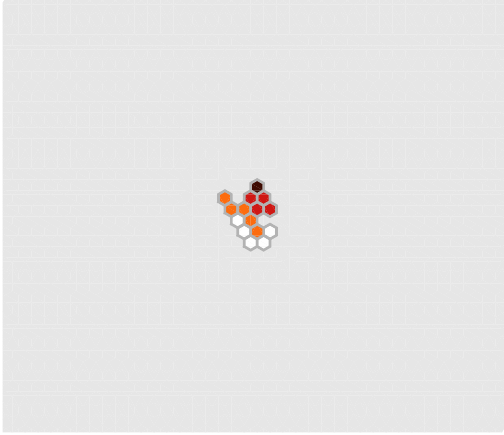}
  \includegraphics[width=0.24\textwidth]{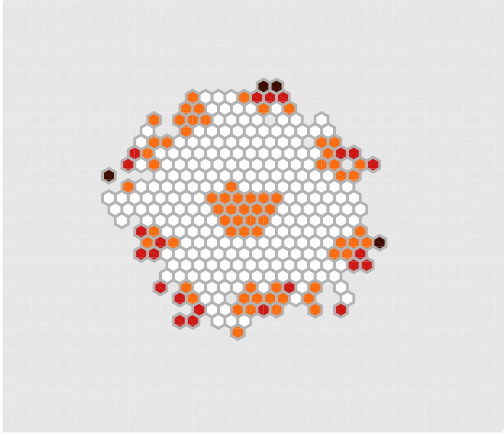}
  \includegraphics[width=0.24\textwidth]{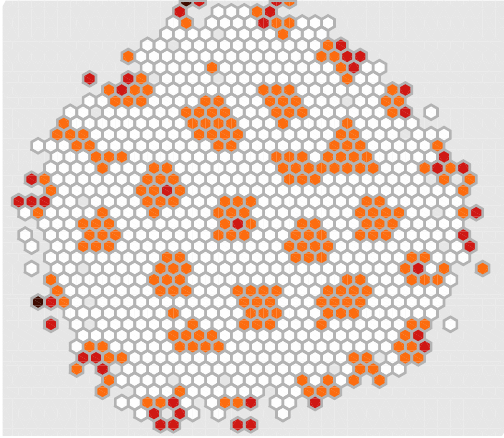}
  \includegraphics[width=0.24\textwidth]{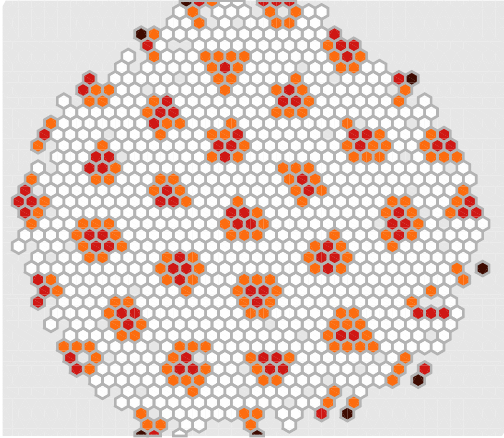}

  \includegraphics[width=0.24\textwidth]{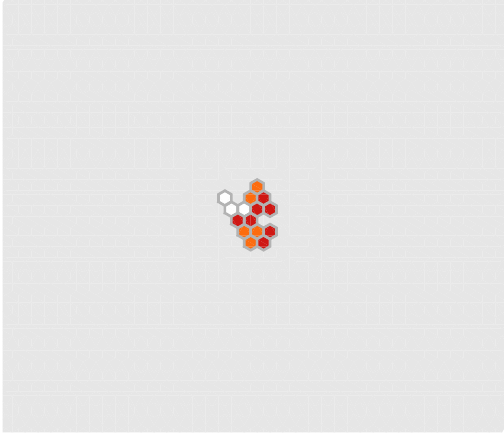}
  \includegraphics[width=0.24\textwidth]{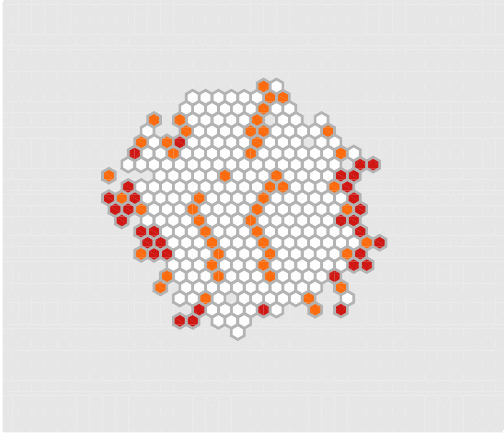}
  \includegraphics[width=0.24\textwidth]{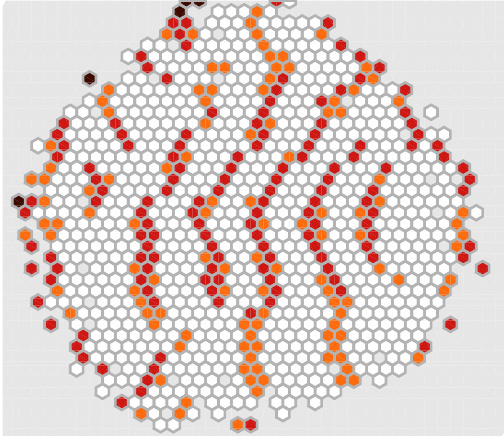}
  \includegraphics[width=0.24\textwidth]{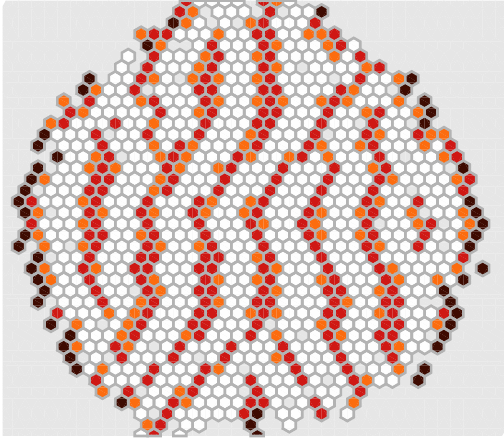}
  \caption{\textit{Top row:} development of spots by differential
    weighting. The parameters are as in Figure~\ref{fig:dynamicRDME1},
    but with $[w_a,q_a,w_b,q_b] = [1,0.001,0.06,0.06]$. \textit{Bottom
      row:} the effect of polarized protrusions. Here the protrusions
    stretch horizontally, $\theta = \pm \pi$, at an angular width
    $[0,\pi/20]$ and a length of 5 cell radii. The parameters are
    again as in Figure~\ref{fig:dynamicRDME1} but with
    $[w_a,q_a,w_b,q_b] = [1,0.001,0.2,0.15]$.}
  \label{fig:dynamicRDME2}
\end{figure}

The coupled inner-outer algorithm can be be understood as a highly
detailed simulation on a very fine mesh covering the whole cell
population, see Figure~\ref{fig:inner_outer}. In
Figure~\ref{fig:dynamicRDME1} we visualize results from the full
Notch-Delta-Reporter model \eqref{eq:NDR}, interpreted in a spatial
stochastic sense as explained in \S\ref{subsec:RDME}, and simulated
together with the growth process as described in
\S\ref{subsec:DLCM}. \review{A small extension to the model was made
  here in that proliferating cells randomly share Notch and Delta in
  between each other, thus adding some noise to the overall dynamics.}
The model itself exhibits a range of intriguing patterns as discussed
in \cite{delta_notch_report}. Two such examples are further
investigated in Figure~\ref{fig:dynamicRDME2}.

%**************************************************************************

\section{Conclusions}

The main focus of this paper has been to investigate the feasibility
of a two-level RDME-DLCM approach. We choose the RDME description of a
single cell as a gold standard modeling approach. This is a detailed,
flexible, yet also comparably effective simulation methodology. At the
cell population level, the related DLCM-method was used and the two
layers of description were coupled together with relative ease.

The main reason this combination is convenient is the fact that both
layers take place in continuous time and can be simulated by
Gillespie-style event-driven algorithms. We also point out that the
overall method combination is promising from the point of view of
deriving approximate simulation algorithms, as for example shown in
detail for the RDME framework in \cite{jsdevarsplit}. A concrete
example is that, for practical reasons in the implementation, we had
to discretize time for the cell-to-cell signaling process of the
model, cf.~\eqref{eq:dtau}. Although not discussed here one can expect
that this method has a strong error of order $O(d\tau^{1/2})$
\cite{jsdesplit}. Since we selected a quite conservative time-step in
\eqref{eq:dtau}, we believe that our implementation is a bit
inefficient in that the time-step restriction is too restrictive given
the accuracy demands of the application at hand. This is an issue
which could clearly be of interest to target in future research
towards faster algorithms. Other related ideas are
deterministic-stochastic hybrid algorithms
\cite{jsdevarsplit,haseltine_HSSA,hybridRD} and, more generally,
multiscale solvers based on ideas from stochastic homogenization
techniques \cite{nestedSSA,smalltimesteps,slowSSA,
  slow_tau}. \review{At the DLCM-layer, the computational bottleneck
  lies in factorizing the Laplacian operator. Real savings in
  computing time here can be expected from employing traditional
  multigrid techniques~\cite{reviewAMG,MULTIGRID_book}.}

% *** table with approx timings
%
% \hline
%  DLCM & 1 minute
%  \hline
% ODE & 5 minutes
% SSA & 60 minutes
% RDME & 24 hours
% \hline

Lastly but not the least, the computational framework described
clearly opens up for many interesting applications where the emerging
cell-population behavior of detailed whole-cell models is to be
approached.

\subsection{Availability and reproducibility}
\label{subsec:reproducibility}

The computational results can be reproduced within the upcoming
release 1.4 of the URDME open-source simulation framework
\cite{URDMEpaper}, available for download at \url{www.urdme.org}.

%**************************************************************************

\section*{Acknowledgment}

Zena Hadjivasiliou kindly and patiently detailed several aspects of
the running model used throughout this paper
\cite{delta_notch_report}.

%**************************************************************************

\newcommand{\doi}[1]{\href{http://dx.doi.org/#1}{doi:#1}}
\newcommand{\available}[1]{Available at \url{#1}}
\newcommand{\availablet}[2]{Available at \href{#1}{#2}}

%\bibliographystyle{abbrvnat}
%\bibliography{../../../stefan}

\providecommand{\noopsort}[1]{} \providecommand{\doi}[1]{\texttt{doi:#1}}
  \providecommand{\available}[1]{Available at \texttt{#1}}
  \providecommand{\availablet}[2]{Available at \texttt{#2}}
\begin{thebibliography}{33}
\providecommand{\natexlab}[1]{#1}
\providecommand{\url}[1]{\texttt{#1}}
\expandafter\ifx\csname urlstyle\endcsname\relax
  \providecommand{\doi}[1]{doi: #1}\else
  \providecommand{\doi}{doi: \begingroup \urlstyle{rm}\Url}\fi

\bibitem[Andrews and Bray(2004)]{Andrews_single_molecule}
S.~S. Andrews and D.~Bray.
\newblock Stochastic simulation of chemical reactions with spatial resolution
  and single molecule detail.
\newblock \emph{Phys.~Biol.}, 1\penalty0 (3):\penalty0 137--151, 2004.
\newblock \doi{10.1088/1478-3967/1/3/001}.

\bibitem[Artavanis-Tsakonas et~al.(1999)Artavanis-Tsakonas, Rand, and
  Lake]{notch_review}
S.~Artavanis-Tsakonas, M.~D. Rand, and R.~J. Lake.
\newblock Notch signaling: Cell fate control and signal integration in
  development.
\newblock \emph{Science}, 284\penalty0 (5415):\penalty0 770--776, 1999.
\newblock \doi{10.1126/science.284.5415.770}.

\bibitem[Atwell et~al.(2015)Atwell, Qin, Gavaghan, Kugler, Hubbard, and
  Osborne]{Atwell:2015:MLM}
K.~Atwell, Z.~Qin, D.~Gavaghan, H.~Kugler, E.~J.~A. Hubbard, and J.~M. Osborne.
\newblock Mechano-logical model of {{\it {C}.~elegans}} germ line suggests
  feedback on the cell cycle.
\newblock \emph{Development}, 142\penalty0 (22):\penalty0 3902, 2015.
\newblock \doi{10.1242/dev.126359}.

\bibitem[Barkai and Leibler(2000)]{circadian}
N.~Barkai and S.~Leibler.
\newblock Circadian clocks limited by noise.
\newblock \emph{Nature}, 403:\penalty0 267--268, 2000.
\newblock \doi{10.1038/35002258}.

\bibitem[Cao and Petzold(2008)]{slow_tau}
Y.~Cao and L.~R. Petzold.
\newblock Slow scale tau-leaping method.
\newblock \emph{Comput.~Meth.~Appl.~Mech.~Engin.}, 197\penalty0 (43):\penalty0
  3472--3479, 2008.
\newblock \doi{10.1016/j.cma.2008.02.024}.

\bibitem[Cao et~al.(2005{\natexlab{a}})Cao, Gillespie, and Petzold]{slowSSA}
Y.~Cao, D.~T. Gillespie, and L.~R. Petzold.
\newblock The slow-scale stochastic simulation algorithm.
\newblock \emph{J.~Chem.~Phys.}, 122\penalty0 (1):\penalty0 014116,
  2005{\natexlab{a}}.
\newblock \doi{10.1063/1.1824902}.

\bibitem[Cao et~al.(2005{\natexlab{b}})Cao, Gillespie, and
  Petzold]{smalltimesteps}
Y.~Cao, D.~T. Gillespie, and L.~R. Petzold.
\newblock Multiscale stochastic simulation algorithm with stochastic partial
  equilibrium assumption for chemically reacting systems.
\newblock \emph{J.~Comput.~Phys.}, 206:\penalty0 395--411, 2005{\natexlab{b}}.
\newblock \doi{10.1016/j.jcp.2004.12.014}.

\bibitem[Chevallier and Engblom(2018)]{jsdevarsplit}
A.~Chevallier and S.~Engblom.
\newblock Pathwise error bounds in multiscale variable splitting methods for
  spatial stochastic kinetics.
\newblock \emph{SIAM J.~Numer.~Anal.}, 56\penalty0 (1):\penalty0 469--498,
  2018.
\newblock \doi{10.1137/16M1083086}.

\bibitem[Cohen et~al.(2010)Cohen, Georgiou, Stevenson, Miodownik, and
  Baum]{filopodia}
M.~Cohen, M.~Georgiou, N.~L. Stevenson, M.~Miodownik, and B.~Baum.
\newblock Dynamic filopodia transmit intermittent delta-notch signaling to
  drive pattern refinement during lateral inhibition.
\newblock \emph{Dev.~Cell}, 19\penalty0 (1):\penalty0 78--89, 2010.
\newblock \doi{10.1016/j.devcel.2010.06.006}.

\bibitem[Collier et~al.(1996)Collier, Monk, Maini, and Lewis]{delta_notch}
J.~R. Collier, N.~A. Monk, P.~K. Maini, and J.~H. Lewis.
\newblock Pattern formation by lateral inhibition with feedback: a mathematical
  model of {D}elta-{N}otch intercellular signalling.
\newblock \emph{J.~Theor.~Biol.}, 183\penalty0 (4):\penalty0 429--446, 1996.
\newblock \doi{10.1006/jtbi.1996.0233}.

\bibitem[Dobrzy{\'n}ski et~al.(2007)Dobrzy{\'n}ski, Rodr{\'\i}guez, Kaandorp,
  and Blom]{Dobrzynski}
M.~Dobrzy{\'n}ski, J.~V. Rodr{\'\i}guez, J.~A. Kaandorp, and J.~G. Blom.
\newblock Computational methods for diffusion-influenced biochemical reactions.
\newblock \emph{Bioinformatics}, 23\penalty0 (15):\penalty0 1969--1977, 2007.
\newblock \doi{10.1093/bioinformatics/btm278}.

\bibitem[Drawert et~al.(2012)Drawert, Engblom, and Hellander]{URDMEpaper}
B.~Drawert, S.~Engblom, and A.~Hellander.
\newblock {URDME}: a modular framework for stochastic simulation of
  reaction-transport processes in complex geometries.
\newblock \emph{BMC Syst.~Biol.}, 6\penalty0 (76):\penalty0 1--17, 2012.
\newblock \doi{10.1186/1752-0509-6-76}.

\bibitem[E et~al.(2005)E, Liu, and Vanden-Eijnden]{nestedSSA}
W.~E, D.~Liu, and E.~Vanden-Eijnden.
\newblock Nested stochastic simulation algorithm for chemical kinetic systems
  with disparate rates.
\newblock \emph{J.~Chem.~Phys.}, 123\penalty0 (19):\penalty0 194107, 2005.
\newblock \doi{10.1063/1.2109987}.

\bibitem[Engblom(2015)]{jsdesplit}
S.~Engblom.
\newblock Strong convergence for split-step methods in stochastic jump
  kinetics.
\newblock \emph{SIAM J.~Numer.~Anal.}, 53\penalty0 (6):\penalty0 2655--2676,
  2015.
\newblock \doi{10.1137/141000841}.

\bibitem[Engblom et~al.(2018)Engblom, Wilson, and Baker]{laplace_cellmech}
S.~Engblom, D.~Wilson, and R.~Baker.
\newblock Scalable population-level modeling of biological cells incorporating
  mechanics and kinetics in continuous time, 2018.
\newblock Accepted for publication in {\it Roy.~Soc.~Open Sci.}
  \available{https://arxiv.org/abs/1706.03375}.

\bibitem[{\noopsort{Engblom20093}}{S.~Engblom and L.~Ferm and A.~Hellander and
  P.~L{\"o}tstedt}(2009)]{master_spatial}
{\noopsort{Engblom20093}}{S.~Engblom and L.~Ferm and A.~Hellander and
  P.~L{\"o}tstedt}.
\newblock Simulation of stochastic reaction-diffusion processes on unstructured
  meshes.
\newblock \emph{SIAM J.~Sci.~Comput.}, 31\penalty0 (3):\penalty0 1774--1797,
  2009.
\newblock \doi{10.1137/080721388}.

\bibitem[Fange and Elf(2006)]{NSM}
D.~Fange and J.~Elf.
\newblock Noise-induced {M}in phenotypes in \textit{E.~coli}.
\newblock \emph{PLoS Comput.~Biol.}, 2\penalty0 (6):\penalty0 637--648, 2006.
\newblock \doi{10.1371/journal.pcbi.0020080}.

\bibitem[Gardiner(2004)]{Gardiner}
C.~W. Gardiner.
\newblock \emph{Handbook of Stochastic Methods}.
\newblock Springer Series in Synergetics. Springer, Berlin, 3rd edition, 2004.

\bibitem[Hadjivasiliou et~al.(2016)Hadjivasiliou, Hunter, and
  Baum]{delta_notch_report}
Z.~Hadjivasiliou, G.~L. Hunter, and B.~Baum.
\newblock A new mechanism for spatial pattern formation via lateral and
  protrusion-mediated lateral signalling.
\newblock \emph{J.~R.~Soc.~Interface}, 13\penalty0 (124):\penalty0 1--10, 2016.
\newblock \doi{10.1098/rsif.2016.0484}.

\bibitem[Haseltine and Rawlings(2002)]{haseltine_HSSA}
E.~L. Haseltine and J.~B. Rawlings.
\newblock Approximate simulation of coupled fast and slow reactions for
  stochastic chemical kinetics.
\newblock \emph{J.~Chem.~Phys.}, 117\penalty0 (15):\penalty0 6959--6969, 2002.
\newblock \doi{10.1063/1.1505860}.

\bibitem[Lestas et~al.(2010)Lestas, Vinnicombe, and
  Paulsson]{fluctuation_limits}
I.~Lestas, G.~Vinnicombe, and J.~Paulsson.
\newblock Fundamental limits on the suppression of molecular fluctuations.
\newblock \emph{Nature}, 467\penalty0 (7312):\penalty0 174--178, 2010.
\newblock \doi{10.1038/nature09333}.

\bibitem[Lo et~al.(2016)Lo, Zheng, and Nie]{hybridRD}
W.-C. Lo, L.~Zheng, and Q.~Nie.
\newblock A hybrid continuous-discrete method for stochastic reaction-diffusion
  processes.
\newblock \emph{Roy.~Soc.~Open Sci.}, 3\penalty0 (9), 2016.
\newblock \doi{10.1098/rsos.160485}.

\bibitem[Naumov et~al.(2011)Naumov, Hoekstra, and Sloot]{Naumov:2011:CAM}
L.~Naumov, A.~Hoekstra, and P.~Sloot.
\newblock Cellular automata models of tumour natural shrinkage.
\newblock \emph{Physica A}, 390\penalty0 (12):\penalty0 2283--2290, 2011.
\newblock \doi{10.1016/j.physa.2011.02.006}.

\bibitem[Puchalka and Kierzek(2004)]{bridging_the_gap}
J.~Puchalka and A.~M. Kierzek.
\newblock Bridging the gap between stochastic and deterministic regimes in the
  kinetic simulations of the biochemical reaction networks.
\newblock \emph{Biophys.~J}, 86\penalty0 (3):\penalty0 1357--1372, 2004.
\newblock \doi{10.1016/S0006-3495(04)74207-1}.

\bibitem[Raj and van Oudenaarden(2008)]{stochgeneexpression}
A.~Raj and A.~van Oudenaarden.
\newblock Nature, nurture, or chance: Stochastic gene expression and its
  consequences.
\newblock \emph{Cell}, 135\penalty0 (2):\penalty0 216--226, 2008.
\newblock \doi{10.1016/j.cell.2008.09.050}.

\bibitem[Robertson-Tessi et~al.(2015)Robertson-Tessi, Gillies, Gatenby, and
  Anderson]{Robertson:2015:IOM}
M.~Robertson-Tessi, R.~J. Gillies, R.~A. Gatenby, and A.~R. Anderson.
\newblock Impact of metabolic heterogeneity on tumor growth, invasion, and
  treatment outcomes.
\newblock \emph{Cancer Res.}, 75\penalty0 (8):\penalty0 1567--1579, 2015.
\newblock \doi{10.1158/0008-5472.CAN-14-1428}.

\bibitem[Shimoni et~al.(2011)Shimoni, Nudelman, Hayot, and
  Sealfon]{multiscaleAgentCell}
Y.~Shimoni, G.~Nudelman, F.~Hayot, and S.~C. Sealfon.
\newblock Multi-scale stochastic simulation of diffusion-coupled agents and its
  application to cell culture simulation.
\newblock \emph{PLoS ONE}, 6\penalty0 (12):\penalty0 1--9, 2011.
\newblock \doi{10.1371/journal.pone.0029298}.

\bibitem[Sprinzak et~al.(2011)Sprinzak, Lakhanpal, LeBon, Garcia-Ojalvo, and
  Elowitz]{mutualNotch}
D.~Sprinzak, A.~Lakhanpal, L.~LeBon, J.~Garcia-Ojalvo, and M.~B. Elowitz.
\newblock Mutual inactivation of notch receptors and ligands facilitates
  developmental patterning.
\newblock \emph{J.~R.~Soc.~Interface}, 7\penalty0 (6):\penalty0 1--11, 06 2011.
\newblock \doi{10.1371/journal.pcbi.1002069}.

\bibitem[St{\"u}ben(2001)]{reviewAMG}
K.~St{\"u}ben.
\newblock A review of algebraic multigrid.
\newblock \emph{J.~Comput.~Appl.~Math.}, 128\penalty0 (1--2):\penalty0
  281--309, 2001.
\newblock Numerical Analysis 2000. Vol. VII: Partial Differential Equations,
  \doi{10.1016/S0377-0427(00)00516-1}.

\bibitem[Trotter et~al.(2001)Trotter, Oosterlee, and
  Sh{\"u}ller]{MULTIGRID_book}
U.~Trotter, C.~W. Oosterlee, and A.~Sh{\"u}ller.
\newblock \emph{Multigrid}.
\newblock Academic Press, 2001.

\bibitem[Turing(1952)]{morphogenesis}
A.~M. Turing.
\newblock The chemical basis of {M}orphogenesis.
\newblock \emph{Phil.~Trans.~R.~Soc.~Lond.~B}, 237\penalty0 (641):\penalty0
  37--72, 1952.
\newblock \doi{10.1098/rstb.1952.0012}.

\bibitem[Vermolen and Gefen(2013)]{Vermolen:2013:ASC}
F.~J. Vermolen and A.~Gefen.
\newblock A semi-stochastic cell-based model for in vitro infected `wound'
  healing through motility reduction: A simulation study.
\newblock \emph{J.~Theor.~Biol.}, 318:\penalty0 68--80, 2013.
\newblock \doi{10.1016/j.jtbi.2012.11.007}.

\bibitem[Ziraldo et~al.(2013)Ziraldo, Mi, An, and Vodovotz]{Ziraldo:2013:CMO}
C.~Ziraldo, Q.~Mi, G.~An, and Y.~Vodovotz.
\newblock Computational modeling of inflammation and wound healing.
\newblock \emph{Adv.~Wound Care}, 2\penalty0 (9):\penalty0 527--537, 2013.
\newblock \doi{10.1089/wound.2012.0416}.

\end{thebibliography}

\providecommand{\noopsort}[1]{} \providecommand{\doi}[1]{\texttt{doi:#1}}
  \providecommand{\available}[1]{Available at \texttt{#1}}
  \providecommand{\availablet}[2]{Available at \texttt{#2}}

%**************************************************************************

\appendix

\section{Simulation algorithms}
\label{app:algorithms}

\begin{algorithm}{\small
  \caption{The Next subvolume method (NSM).}
  \label{alg:NSM}
  \begin{algorithmic}

    \STATE{\textit{Initialize:} Distribute the molecular species over
      all voxels and compute the sums $\sigma_i^r$ and $\sigma_i^d$ of
      all reaction- and diffusion rates, $i = 1,\ldots,\Nvoxels$.
      Sample the waiting time $\tau_i$ until the next event in voxel
      $i$ by drawing an exponentially distributed random variable of
      intensity $\sigma_i^r+\sigma_i^d$, and store all these waiting
      times in a heap.}

    \WHILE{$t < T$}

    \STATE{Select the next voxel $i$ where an event takes place by
      extracting the minimum $\tau_i$ from the top of the heap.}

    \STATE{Determine if the event is a reaction- or a diffusion event
      by inverse sampling.}

    \IF{Reaction event}
  
    \STATE{Determine the reaction channel that fires. This is done by
      Gillespie's Direct method.}

    \STATE{Update the state accordingly.}

    \STATE{Update the rates $\sigma_i^r$ and $\sigma_i^d$ to take the
      updated state into account.}

    \ELSE[Diffusion event]

    \STATE{Use Gillespie's Direct method to determine which species
      diffuses and to which neighboring voxel $j$.}

    \STATE{Update the state accordingly.}
  
    \STATE{Update the reaction- and diffusion rates in voxels $i$ and
      $j$.}
  
    \ENDIF
  
    \STATE{ Set $t = \tau_i$. Compute a new waiting time $\tau_i$ by
      drawing a new random number and adding it to the heap.}
  
    \ENDWHILE
  \end{algorithmic}}
\end{algorithm}

\begin{algorithm}{\small
  \caption{DLCM simulation by Gillespie's Direct method.}
  \label{alg:DLCM}
  \begin{algorithmic}

    \STATE{\textit{Initialize:} Given a state $(u_i)$,
      $u_i \in \{0,1,2\}$ over the mesh of voxels $(v_i)$,
      $i = 1,\ldots,\Nvoxels$.}

    \WHILE{$t < T$}

    \STATE{Solve for the cellular pressure $(p_i)$,
      \eqref{eq:dlaplace}--\eqref{eq:hom_dirichlet}. Compute all
      movement rates $R_{ij}$, \eqref{eq:rate}.}

    \STATE{Determine the rates for any other processes taking place in
      the model, e.g., proliferation and death events, or active
      migration.}

    \STATE{Sample the next event time $\tau$ and the kind of event
      using Gillespie's procedure.}

    \STATE{Update the state of all cells with respect to any other
      continuous-time processes taking place in $[t,\tau)$, e.g.,
      intracellular kinetics or cell-to-cell communication.}

    \STATE{Update the state $(u_i)$ by executing the state transition
      associated with the determined event and set $t = \tau$.}

    \ENDWHILE
  \end{algorithmic}}
\end{algorithm}

\end{document}